\documentclass [aps,floatfix,subeqn,twocolumn,amsmath]{revtex4}
\usepackage{amssymb}
\usepackage{epsfig}
\usepackage{graphicx}
\usepackage{natbib}
\usepackage{picinpar}

\begin{document}

\draft

\title{Normal modes of a quasi-one-dimensional multi-chain complex plasma}

\author{{G. Piacente$^1$, F.~M. Peeters$^1$ and J.~J. Betouras$^{1,2}$}}

\address{$^1$Department of Physics, University of Antwerp (Campus Drie
Eiken), B-2610 Antwerpen, Belgium \\
$^2$ Max Planck Institute for the Physics of Complex Systems,
Noethnitzer Strasse 38, Dresden 01187, Germany}

\date{\today}

\begin{abstract}
We studied equally charged particles, suspended in a complex
plasma, which move in a plane and interact with a screened Coulomb
potential (Yukawa type) and with an additional external confining
parabolic potential in one direction, that makes the system
quasi-one-dimensional (Q1D). The normal modes of the system are
studied in the presence of dissipation. We also investigated how a
perpendicular magnetic field couples the phonon modes with each
other. Two different ways of exciting the normal modes are
discussed: 1) a uniform excitation of the Q1D lattice, and 2) a
local forced excitation of the system in which one particle is
driven by e.g. a laser. Our results are in very good agreement
with recent experimental findings on a finite single chain system
(Phys. Rev. Lett. {\bf 91}, 255003 (2003)). Predictions are made
for the normal modes of multi-chain structures in the presence of
damping.
\end{abstract}


\maketitle

\section{Introduction}

Since the first experimental observations of the formation of
Coulomb quasi-lattices \cite{chu} involving highly charged dust
particulate in 1994 \cite{thomas,hayashi,melzer}, the research
field of complex plasmas has seen a sustained growth. Complex
plasmas consist of micrometer-sized (\emph{"dust"}) particles
immersed in a gaseous plasma background. Dust particles typically
acquire a negative charge of several thousand elementary charges,
so they interact with each other through their strong
electrostatic repulsion. In the experiment, particles are trapped
in a horizontal layer by a shallow parabolic well, due to two
electrodes, and can be suspended in a sheath above the electrodes,
where the gravity force is balanced by the electrostatic force.
When the electrostatic energy of neighboring particles exceeds the
thermal energy by an amount $\Gamma$, the particles arrange
themselves in regular, solid-like structures, i.e. Wigner crystals
\cite{wigner}. The mutual Coulomb repulsion of the dust grains is
partly screened by the polarization of the surrounding plasma
particles, mostly by the gas ions which represent the major
species in the sheath. Therefore, the average interparticle
potential can be well represented by a Yukawa
(Debye-H$\ddot{u}$ckel) potential \cite{ikezi}.

Complex plasmas provide a new system for the study of classical
crystalline and liquid dynamics and the melting processes. For
particle size of the order of $\mu$m, the dynamical behavior can
be monitored directly with the use of optical microscopes
\cite{chu}.

In the present work we study thoroughly the normal modes of a
classical quasi-one-dimensional (Q1D) multi-chain complex plasma.
Such a Q1D system was recently realized experimentally by giving a
proper shape to the electrodes \cite{goree,homann}. Experimentally
many other quasi-one-dimensional or strictly one-dimensional
systems have been realized over the years. Colloidal particles,
suspended in aqueous solution, can be trapped in a potential well
created by two counterpropagating laser beams which form a one
dimensional coupled array \cite{tatarkova}. A Coulomb chain
confined in a storage ring \cite{birkl}, as well as ordered
electrons on microchannels filled by liquid Helium \cite{glasson}
are other examples of Q1D classical systems. The latter is one of
the candidates to be used for quantum computing
\cite{cirac,dykman}. On the atomic scale a chain-like system can
be found in compounds such as ${Hg}_{3-\delta}AsF_6$ \cite{brown}
and in low dimensional systems formed on surfaces \cite{segovia}.
A one dimensional chain of gas atoms adsorbed by carbon nanotubes
has been produced in laboratory and its phonon spectrum has been
calculated theoretically assuming a Lennard-Jones interaction
potential \cite{cvitas,siber}.

The classical model, we propose in the present paper, reveals a
non trivial phase diagram at zero temperature and allows to
calculate dispersion relations for the normal modes, which can be
directly investigated experimentally. Several generic aspects of
the model were investigated recently \cite{piacente}. Here, the
main focus is on the normal modes of the system and how they
depend on different physical situations, e.g. frictional forces
and the way they are excited. We will make connection with recent
experimental works \cite{goree, goree2}.

The structure of the paper is as follows: we first give in Sec. II
an overview of the model, stressing the ground state properties.
In Sec. III we present what has been known so far for the normal
modes of the system, adding more results for clarifications, and
then we turn to the new results with respect to the presence of
the gas drag (friction) with or without an applied magnetic field.
Finally, we turn our attention to the forced oscillations in Sec.
V, where we discuss recent experiments on normal modes in single
chain systems. Prediction for multi-chain configurations are
presented in Sec VI. Finally we conclude in Sec. VII.

\section{Model and general properties}

We consider a  system of equally charged particles with
coordinates $\vec{r}_i=(x_i,y_i)$ moving in a plane and
interacting with each other through a Yukawa-type potential (the
screening length $\lambda$ is an external parameter which is
measured in the experiment \cite{konopka}) and are confined by a
parabolic potential in the $y$-direction. The dimensionless
Hamiltonian of the system is given by:
\begin{equation}
H'=\sum_{i \neq j} \frac{exp(-\kappa |\vec{r'_i}-\vec{r'_j}|)}
{|\vec{r'_i}-\vec{r'_j}|} + \sum_i {y'_i}^2 ,
\end{equation}
where $H^{\prime }=H/E_{0}$, $\kappa =r_{0}/\lambda $ and
$\overrightarrow{r}^{\prime }=\overrightarrow{r}/r_{0}$, with
$r_{0}=(2q^2/m\varepsilon \omega _{0}^{2})^{1/3}$ as unit of
length and $E_{0}=(m\omega _{0}^{2}q^{4}/2\varepsilon ^{2})^{1/3}$
as unit of energy; $m$ and $q$ are the mass and the charge of the
particles, respectively, $\epsilon$ is the dielectric constant of
the medium the particles are moving in and $\omega_0$ measures the
strength of the confining potential. The dimensionless time is
defined as $t'=\omega_0t$. Finally, it is possible to define a
dimensionless temperature as $T^{\prime }=T/T_{0}$  with $%
T_{0}=E_0/k_{B}=(m\omega _{0}^{2}q^{4}/2\varepsilon
^{2})^{1/3}k_{B}^{-1}$.

In our previous work \cite{piacente} we investigated the ground
state and the melting of this Q1D system. We summarize here the
main results, which we will need in the next sections. At $T=0$
the particles crystallize in a chain-like crystal structure, with
a linear density equally distributed among the chains. In the case
of multiple chains, if $a$ is the separation between two
neighboring particles in the same chain, the chains are staggered
by $a/2$ in the $x$-direction, because this arrangement minimizes
the electrostatic repulsion. The results for the ground state
configuration are summarized in the phase diagram depicted in Fig.
1.

\begin{figure}
\begin{center}
\includegraphics[width=7.5cm]{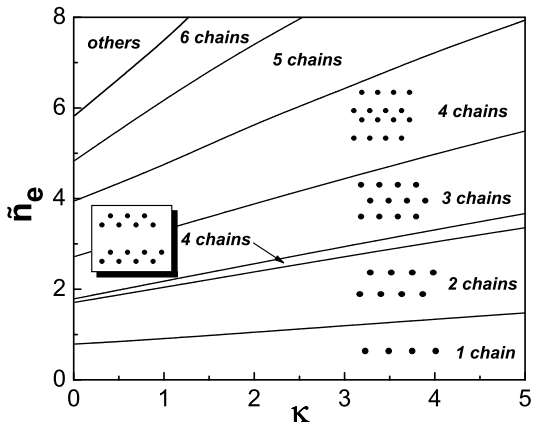}
\caption{The $T=0$ structural phase diagram as a function of the
inverse screening length $\kappa$ and the density
${\widetilde{n}}_e$. The plotted quantities are dimensionless, as
it will be for all the figures in the paper.}
\end{center}
\end{figure}

The dimensionless linear density is defined as $\
\widetilde{n}_{e}=\nu r_{0}/a$, where $\nu$ is the number of
chains. For low densities the particles crystallize in a single
chain; with increasing density a continuous transition
(''zig-zag'' \cite{zigzag}) occurs and the single chain splits
into two chains. Further increasing the density we found the
remarkable behavior that the four-chain structure is stabilized
before the three-chain structure. The 2 $\rightarrow$ 4 chain
transition occurs through a ``zig-zag'' transition of each of the
chains accompanied by a shift of $a/4$ along $x$. This 4 chain
configuration has a relatively small stability range after which
it transits to a 3 chain configuration through a discontinuous,
i.e. first order phase transition. For higher values of the
density, the 4 chain configuration attains again the lowest
energy. A further increase of $\widetilde{n}_{e}$ will lead to
more chains, that is 6, 7 and so on. The structural transitions
are discontinuous, i.e. first order, except for the 1
$\rightarrow$ 2 transition.

Raising the temperature the ordered structure melts. We studied in
detail the melting for such a kind of system in Ref. [18]. Due to
the anisotropy in the two directions, a different behavior of the
system in $x$ and $y$ directions was found. Two different melting
temperatures, $T_x$ and $T_y$, can be assigned. The main features
of the transition from the ordered state to the liquid state are:
i) a reentrant behavior as a function of density; ii) a region in
density for which the system melts first in the unconfined
direction and then in the confined one: this regime resembles the
findings of Ref. \cite{radzihovsky} in the floating solid regime;
iii) the reentrant melting occurs near the structural transition
points. For the nearly Coulomb limit ($\kappa$=0.01) there is no
evidence of anisotropic melting, that is the system behaves more
isotropically. Furthermore, the Coulomb system has a melting
temperature which is on average 15-20 \% higher than for the
screened Coulomb inter-particle interaction with $\kappa = 1$. On
the other hand for higher value of $\kappa$, the system behaves
more anisotropically and the difference between $T_x$ and $T_y$ is
enhanced. In the case $\kappa=3$ the melting temperature is on
average 10-15 \% lower than for the case $\kappa = 1$.

\section{Normal modes}
In the present paper we are interested in the normal modes of the
chain structures, and in particular how these modes are modified
in the presence of frictional forces. We will consider: 1) the
uniformly damped motion of a normal mode, and 2) the damped
propagation of a local forced oscillation of a single particle.
For these purposes we review shortly the normal modes in the
absence of friction.

\subsection{Dispersion relations in the absence of friction}

In the absence of drag due to the ion gas and exploiting the
standard harmonic approximation, the equations of motion for small
oscillations about the lattice equilibrium positions in
dimensionless units are in the single chain case:

\begin{subequations}
\begin{equation}
\begin{split}
\frac{d^{2}{x'_i}}{d{t'^{2}}}&=-\frac{1}{2}\sum_{j}\frac{\partial^{2}{U}}
{\partial{x'_i}\partial{x'_j}}\Bigg{|}_{eq}x'_j \ + \\&-
\frac{1}{2}\sum_{j}\frac{\partial^{2}{U}}
{\partial{x'_i}\partial{y'_j}}\Bigg{|}_{eq} y'_j
\end{split}
\end{equation}
\begin{equation}
\begin{split}
\frac{d^{2}{y'_i}}{d{t'^{2}}}&=-\frac{1}{2}\sum_{j}\frac{\partial^{2}{U}}
{\partial{y'_i}\partial{x'_j}}\Bigg{|}_{eq} x'_j \ + \\&-
\frac{1}{2}\sum_{j}\frac{\partial^{2}{U}}
{\partial{y'_i}\partial{y'_j}} \Bigg{|}_{eq} y'_j \ -\ y'_i
\end{split}
\end{equation}
\end{subequations}

\noindent where $U=\exp(-\kappa
|\vec{r'_i}-\vec{r'_j}|)/{|\vec{r'_i}-\vec{r'_j}|}$ is the
interparticle interaction potential. Considering the translational
invariance of the system along the $x$ direction, we search for
solutions in the form
\begin{equation}
(x'_n,y'_n)\propto\exp{[i(kna-\omega t)]},
\end{equation}
\noindent which results into
\begin{equation}
[(\omega^{2}-\delta_{\beta
y})\delta_{\alpha\beta,ij}-D_{\alpha\beta,ij}]Q_{\beta,j}=0
\end{equation}
where $D_{\alpha\beta,ij}$ is the dynamical matrix, that is the
matrix of the second derivatives of the Yukawa potential,
calculated at the equilibrium configuration; $Q_{\beta,j}$ is the
displacement of particle $j$ from its equilibrium position in the
$\beta$ direction; $(\alpha,\beta)\equiv{(x,y)}$,
$\delta_{\alpha\beta,ij}$, $\delta_{ij}$ and $\delta_{\beta y}$
are unit matrices, in particular $\delta_{\beta y}$ takes into
account the effect of the confining potential. All the frequencies
are measured in unit of $\omega_{0}$.

The number of chains determine the number of particles in each
unit cell and therefore the number of degrees of freedom per unit
cell. So if $l$ is the number of chains there will be $2l$
branches for the normal mode dispersion curves. Note that for
ordinary bidimensional crystals there are 2 acoustical branches
and $2p - 2$ optical branches \cite{ashkroft}, if $p$ is the
number of atomic species in the unit cell.

Solving explicitly Eq. (4) for the single chain configuration we
obtain that the acoustical and optical eigenfrequencies are given
respectively by:

\begin{subequations}
\begin{equation}
\begin{split}
\omega_{ac}(k) = \Big{\{}{\tilde{n}_e}^3 \sum_{j=1}^{\infty}
\frac{\exp(-j\kappa /{\tilde{n}_e})}{j^3} \ \times \\
(2+\frac{2j\kappa} {\tilde{n}_e} + \frac { j^2
{\kappa}^2}{{\tilde{n}_e}^2}) [1-\cos(kaj)] \Big{\}}^{1/2} \ ,
\end{split}
\end{equation}

\begin{equation}
\begin{split}
\omega_{opt} (k)= \Big{\{}1-{\tilde{n}_e}^3 \sum_{j=1}^{\infty}
\frac{\exp(-j \kappa /{\tilde{n}_e})}{j^3} \ \times \\(1 +
\frac{j\kappa}{\widetilde{n}_e} ) [1-\cos(kaj)] \Big{\}}^{1/2} \ ,
\end{split}
\end{equation}
\end{subequations}

\noindent where $k$ is the wave number. Numerical results for the
dispersion relations were presented in Ref. [18].

It is interesting to notice that for the acoustical branch the
dispersion is positive, that is phase and group velocity have the
same sign, while for the optical branch the dispersion is
negative, i.e. the group velocity is negative. Physically the
negative dispersion for the single chain optical branch can be
understood considering that the electrostatic repulsion acts
oppositely to the force of the confining potential and this
reduces the oscillation frequency with increasing $k$.

Another notable feature is the softening of the optical branch,
accompanied by a hardening of the acoustical branch at the values
of $\widetilde{n}_{e}$ and $\kappa$ where the $1 \rightarrow 2$
structural transition is observed (see Fig. 8 of Ref. [18]), which
confirms that $1 \rightarrow 2$ is a continuous transition.

When a magnetic field is applied in the perpendicular direction to
the plane the particles are moving in, the equations of motions
are modified and ${\dot{y}'_i}\omega'_c$ is added to the right
hand side of Eq. (2a) and $-{\dot{x}'_i}\omega'_c$ to the right
hand side of Eq. (2b), where $\vec{\omega_{c}} = q\vec{B}/mc$ is
the cyclotron frequency and $\omega'_{c}=\omega_{c}/\omega_0$. It
is known \cite{vanlee} that in a classical system an external
magnetic field does not alter the statistical properties of the
system and consequently the structural properties should be
insensitive to the magnetic field strength. On the other hand the
character of motion of the particles is altered significantly
because now the $x$ and $y$ motion are coupled. The spectrum of an
infinite bidimensional crystal in a magnetic field was obtained in
Refs. \cite{chaplik,bonsall}. Following Ref. \cite{bonsall}, the
dispersion relation for our Q1D system in the presence of a
perpendicular magnetic field $B$ is obtained from
\begin{equation}
[(\omega^{2}-\delta_{\beta
y})\delta_{\alpha\beta,ij}-D_{\alpha\beta,ij}+
i\omega\omega_{c}\xi_{\alpha\beta}\delta_{ij}]Q_{\beta,j}=0.
\end{equation}

\subsection{Dispersion relations in the presence of friction}

In laboratory experiments of a dusty plasma the particles
experience a frictional drag due mainly to the background neutral
gas as well as ions. This drag has a significant effect on the
dispersion curves of the normal modes. In order to compare
experimental data with theory, it is necessary to develop a
theoretical model in which the structure of the crystal as well as
damping are included as essential elements. This can be easily
done by adding explicitly the friction term in the equations of
motion. For the single chain configuration $\gamma'\dot{x}'_i$
should be added to the left hand side of Eq. (2a) and
$\gamma'\dot{y}'_i$ to the left hand side of Eq. (2b), where
$\gamma'=\gamma/\omega_0$ is the dimensionless frictional drag
coefficient. Similar equations hold naturally in the multi-chain
structures. The equations of motion for the 2 and 3 chain
structures are reported for completeness in Appendix A.

Proceeding as before, in this case the eigenfrequencies are
determined by:
\begin{equation}
[(\omega^{2}-\delta_{\beta
y}+i\gamma\omega)\delta_{\alpha\beta,ij}-D_{\alpha\beta,ij}]Q_{\beta,j}=0.
\end{equation}
For a single chain Eq. (7) gives explicitly:
\begin{subequations}
\begin{equation}
\begin{split}
\omega_{ac}^2+i\gamma\omega_{ac} - {\tilde{n}_e}^3
\sum_{j=1}^{\infty} \frac{\exp(-j\kappa
/{\tilde{n}_e})}{j^3} \ \times \\
 (2+ \frac{2j\kappa} {\tilde{n}_e} +
\frac { j^2 {\kappa}^2}{{\tilde{n}_e}^2}) [1-\cos(k a j)]=0,
\end{split}
\end{equation}
\begin{equation}
\begin{split}
\omega_{opt}^2+i\gamma\omega_{opt}-1+{\tilde{n}_e}^3
\sum_{j=1}^{\infty} \frac{\exp(-j \kappa /{\tilde{n}_e})}{j^3} \
\times \\
(1 + \frac{j\kappa}{\tilde{n}_e} ) [1-\cos(k a j)]=0,
\end{split}
\end{equation}
\end{subequations}

\noindent from which we obtain the solutions:

\begin{subequations}
\begin{equation}
\begin{split}
\omega_{ac}(k) = \Big{\{}{\tilde{n}_e}^3 \sum_{j=1}^{\infty}
\frac{\exp(-j\kappa /{\tilde{n}_e})}{j^3} \ \times&\\
(2+\frac{2j\kappa} {\tilde{n}_e} + \frac { j^2
{\kappa}^2}{{\tilde{n}_e}^2}) [1-\cos(k a
j)]-\frac{\gamma^2}{4}\Big{\}}^{1/2} \ -i\frac{\gamma}{2},
\end{split}
\end{equation}
\begin{equation}
\begin{split}
\omega_{opt} (k)= \Big{\{}1-{\tilde{n}_e}^3 \sum_{j=1}^{\infty}
\frac{\exp(-j \kappa /{\tilde{n}_e})}{j^3} \ \times&\\(1 +
\frac{j\kappa}{\tilde{n}_e} ) [1-\cos(k a
j)]-\frac{\gamma^2}{4}\Big{\}}^{1/2}\ -i\frac{\gamma}{2}.
\end{split}
\end{equation}
\end{subequations}
\noindent The analytical expression for the 2 and 3 chain
eigenfrequencies are reported in Appendix B.

In the limit of small wavenumber $k$, we find for
$\kappa/\widetilde{n}_e \gg 1$ that Eqs (9a) and (9b) reduce
respectively to:
\begin{subequations}
\begin{equation}
\begin{split}
\omega_{ac}(k) = \Big{\{}e^{-\kappa/\widetilde{n}_e}
\frac{{\kappa}^2\widetilde{n}_e} {2} \ \times\\(1-\frac{k^2
a^2}{12}) k^2 a^2 -\frac{\gamma^2}{4}\Big{\}}^{1/2}-\
i\frac{\gamma}{2},
\end{split}
\end{equation}
\begin{equation}
\begin{split}
\omega_{opt}(k) = \Big{\{}1-  e^{-\kappa/\widetilde{n}_e}
\frac{{\widetilde{n}_e}^2 \kappa}{2}\ \times\\(1-\frac{k^2
a^2}{12}) k^{2}{a^{2}}-\frac{\gamma^2}{4}\Big{\}}^{1/2}-\
i\frac{\gamma}{2},
\end{split}
\end{equation}
\end{subequations}
\noindent while for $\kappa/\widetilde{n}_e \ll1$ we find:
\begin{subequations}
\begin{equation}
\begin{split}
\omega_{ac}(k) = \Big{\{}[\frac{3}{2} +
\ln(\frac{\widetilde{n}_e}{\kappa})-\frac{\widetilde{n}_e}{12}\ \times\\
(1+\frac{5\widetilde{n}_e}{12})k^2 a^2] {\widetilde{n}_e}^{3} k^2
a^2-\frac{\gamma^2}{4}\Big{\}}^{1/2} -\ i\frac{\gamma}{2},
\end{split}
\end{equation}
\begin{equation}
\begin{split}
\omega_{opt}(k) = \Big{\{} 1 -[
1+\ln(\frac{\widetilde{n}_e}{\kappa})-\frac{\widetilde{n}_e}{\kappa}\ \times\\
(1+\frac{2\widetilde{n}_e}{\kappa})k^2
a^2]\frac{{\widetilde{n}_e}^3}{2} \frac{k^2
a^2}{12}-\frac{\gamma^2}{4}\Big{\}}^{1/2} -\ i\frac{\gamma}{2}.
\end{split}
\end{equation}
\end{subequations}

\begin{figure}
\begin{center}
\includegraphics[width=7.5cm]{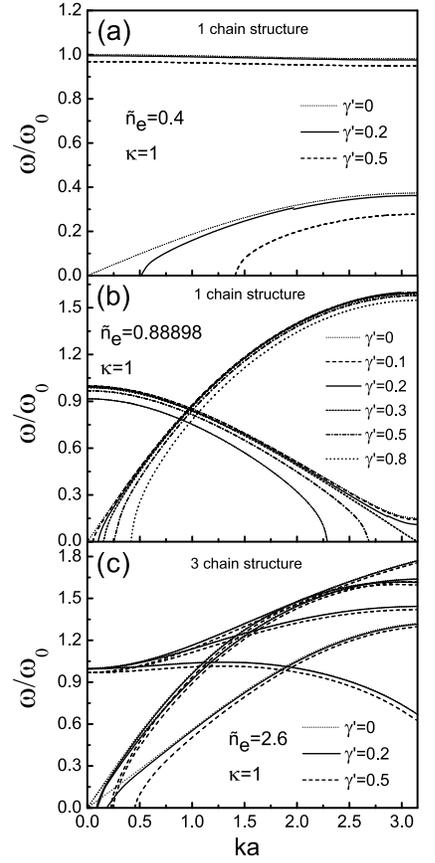}
\caption{Dispersion curves for the normal modes in the presence of
friction for different values of parameters for a one (a), three
(c) chain structure. Dependence of the acoustical and optical
branches in the one chain case on the friction coefficient at the
critical density where the softening of the optical mode is
observed is given in (b).}
\end{center}
\end{figure}

The real part of the frequency corresponds to the oscillatory
motion while the damping in the time domain is given by the
imaginary term $i \gamma/2$. Note that friction also effects value
of the frequencies of normal modes. In Fig. 2 the normal mode
spectra are reported for the different configurations of the
system for different values of $\widetilde{n}_e$, $\kappa$ and
$\gamma'$. We used in our calculations values for the parameters
inferred from the experiment \cite{goree}. The behavior of the
dispersion curves reflects rather closely the case without
damping. Some new features should, however, be stressed: i) the
effect of friction results in general into a reduction of the
frequencies of vibration both for the longitudinal motion and for
the transversal one; ii) for very small values of the wave number
the acoustical vibrations cannot be excited, which implies that
they are overdamped. Such waves can only be excited when
$k>k^*(\widetilde{n}_e,\kappa,\gamma')$; iii) the softening of the
optical mode at the critical density $\widetilde{n}_e^{*}$ for the
transition $1\rightarrow 2$ depends on $\gamma'$, in particular
the presence of friction reduces the value of
$\widetilde{n}_e^{*}$ (see Fig. 2(b)).

\begin{figure}
\begin{center}
\includegraphics[width=7.5cm]{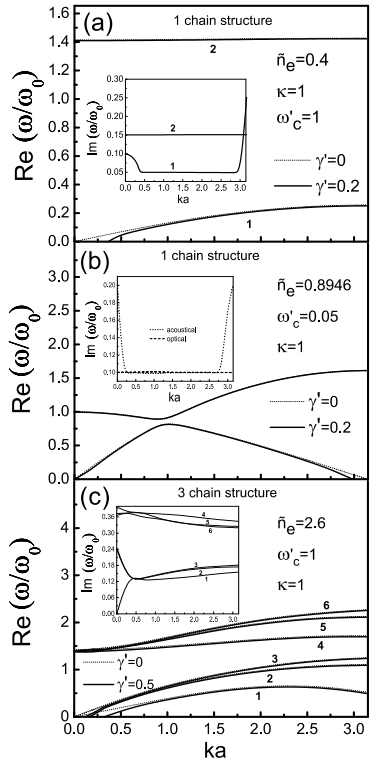}
\caption{The same as Fig. 2 but now a magnetic field of stregth
$\omega_c/\omega_0=1$ is present. The insets depict the damping of
magneto-phonon modes.}
\end{center}
\end{figure}

When we include a magnetic field the damped normal modes are
determined by:

\begin{equation}
\begin{split}
[(\omega^{2}-\delta_{\beta
y}+i\gamma\omega)\delta_{\alpha\beta,ij}+\\-D_{\alpha\beta,ij}+
i\omega\omega_{c}\xi_{\alpha\beta}\delta_{ij}]Q_{\beta,j}=& \ 0.
\end{split}
\end{equation}

\noindent The corresponding dispersion curve for the single and
multi-chain structures are reported in Fig. 3. The behavior of the
curves resembles the case without damping but with an additional
shift in frequency due to friction. Note that in this case it is
no longer possible to obtain the phonon frequencies analytically.
The anticrossing between the two branches in the 1 chain
configuration is still present (see Fig. 3(b)), as in the case
without friction (see Fig. 11 of Ref. \cite{piacente}). It is
remarkable that the cyclotron motion and the friction are coupled
and the magnetic field introduces a dispersion in the imaginary
part of $\omega$ as well. Now, $Im(\omega)$ is no longer constant
as a function of the wavevector as in the case without a magnetic
field. Friction mainly alters the acoustical branches of the
magneto-phonon modes for small wavevectors.

\section{Forced oscillations in a single chain structure}

In the experiment of Refs. \cite{goree, rosenberg, pieper,
homann2, nanomura, liu} the system is set into oscillation by an
external driving force which acts on the system continuously. The
frequency of such a \emph{forced oscillation} is then determined
by the frequency of the driving force and not by the resonant
frequencies. This is the effective situation in the experiments
where particle motions are excited by laser manipulation, which
makes it possible to excite and test the dispersion relations of
certain types of lattice wave
\cite{wang,rosenberg,pieper,homann2}, which are longitudinal waves
and, most recently also, transverse waves were observed
\cite{goree,nanomura,vladimirov}. Laser light exerts a radiation
pressure on the particles with a magnitude proportional to the
laser intensity \cite{liu}. In these cases the frequency is purely
real since the modes are driven. What is observed in the
experiments, is that as the wave propagates it is spatially
damped, which can be interpreted in term of a complex wave number
\cite{wang} $k=k_r+ik_i$. Following this idea and considering that
excitations take place when the driving frequency is close to the
free frequency of the modes, we may neglect to first approximation
for the single chain structure the external force and we looked
for particular solutions of the equations of motion, in the form:
\begin{equation}
(x'_n,y'_n)\propto\exp{[i(k_rna-\omega t)]}\exp{(-k_ina)}
\end{equation}
\noindent as it was done in Ref. \cite{goree} for the theoretical
calculation of the optical branch. This yields for the acoustical
and optical branch, respectively:
\begin{widetext}
\begin{subequations}
\begin{equation}
\omega_{ac}^2+i\gamma\omega_{ac} -
{\tilde{n}_e}^3\sum_{j=1}^{\infty} \frac{\exp(-j\kappa
/{\tilde{n}_e})}{j^3} (2+ \frac{2j\kappa} {\tilde{n}_e} +
\frac{j^2\kappa^2}{{\tilde{n}_e}^2}) [1-\cos(k_r a j) \cosh{(k_i a
j)}+i\sin(k_r a j)\sinh(k_i a j)]=0,
\end{equation}
\begin{equation}
\omega_{opt}^2+i\gamma\omega_{opt}-1+{\tilde{n}_e}^3\sum_{j=1}^{\infty}
\frac{\exp(-j\kappa /{\tilde{n}_e})}{j^3}(1 +
\frac{j\kappa}{\tilde{n}_e} ) [1-\cos(k_r a j) \cosh{(k_i a
j)}+i\sin(k_r a j)\sinh(k_i a j)]=0.
\end{equation}
\end{subequations}
\end{widetext}

\begin{figure}
\begin{center}
\includegraphics[width=7.5cm]{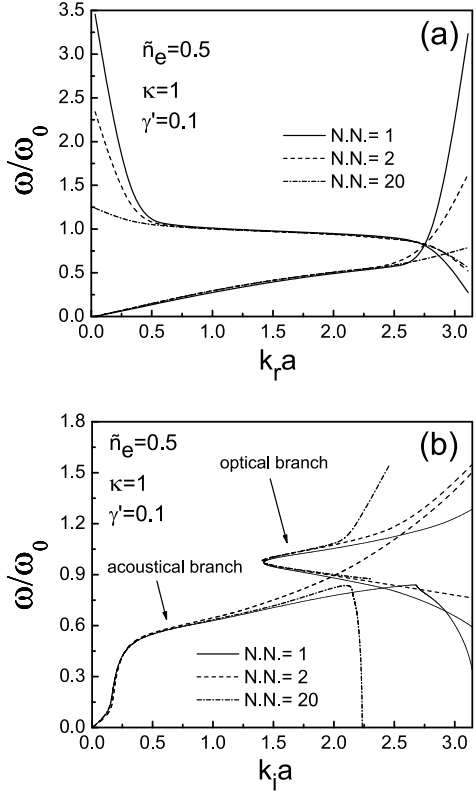}
\caption{Dispersion relations for a single chain structure as a
function of the: (a) real part and (b) imaginary part of the
wavevector. The curves show a strong dependence on N.N., the
number of nearest neighbors included in the calculations, at the
edges of the first Brillouin zone.}
\end{center}
\end{figure}

Requiring the frequency $\omega$ to be real, both equations
generate a system of two non linear equations, for $k_r$ and
$k_i$. The results of the calculation are reported in Fig. 4. This
approach has, however, some limits. First of all, once the laser
acts on a specific particle \cite{goree}, it is no longer possible
to consider all the particle as identical, i.e. the presence of an
external force breaks the symmetry of the system which is taken
into account in Eq. (13) by considering the driven particle as
being $x_{n=0}=0$. Second, the convergence of the series in Eq.
(14a) and Eq. (14b) is no longer guaranteed. The condition that
must be fulfilled in order to have a convergent sum is $k_i a\leq
\kappa/ \tilde{n}_e$. As seen from Fig. 4(b), this condition is
not always satisfied (in the specific case considered in the
picture the condition for convergence is $k_i a \leq 2$). Note
that the dispersion curves depend strongly on the number of terms
considered in the sum at the edges of the first Brillouin zone
(Fig. 4(a)), while it is practically independent of the number of
neighbors considered in the sum in the middle of the first
Brillouin zone. The system of equations arising from Eqs. (14a)
and (14b) is not defined when $ka=0$ and $ka=\pi$, because in this
case the equation for the imaginary part is identically zero. This
clearly shows that inapplicability of this approach to obtain the
phonon spectrum in the presence of friction.

The reason for the divergence of the sums in Eqs. (14a) and (14b)
is a consequence of the fact that the last exponent in Eq. (13)
blows up for negative values of $n$. This would suggest that
alternatively we should look for solutions of the equation of
motion in the form:

\begin{equation}
(x'_n,y'_n)\propto\exp{[i(k_rna-\omega t)]}\exp{(-k_i|n|a)},
\end{equation}
\noindent i.e. damped waves propagating form the location of
external excitation. But in this case in the imaginary part in
Eqs. (14a) and (14b) the hyperbolic sine term is replaced by
$\exp{(-k_i|j|a)}$ and consequently the sum gives zero. As a
result, Eqs. (14a) and (14b) do not have any real solutions for
the phonon frequency and this approach also fails.

 In order to
explain some recent experimental results on the transversal modes
of a finite one dimensional chain \cite{goree}, excited by
striking one particle with two counterpropagating laser beams such
that the effective force acting on the particle is $I_0\sin{\omega
t}$, with $I_0$ the intensity of the beam, we have followed
another approach. We first consider a single finite chain of $N$
particles confined in the $y$ direction. On one of the particles a
time varying force is acting. We studied the small displacements
from the equilibrium configuration of each particle, limiting
ourselves to first neighbor interactions, which is valid for
$ka>1$. The equations of motion for such a system are:
\begin{subequations}
\begin{equation}
\begin{split}
\frac{d^{2}{x'_l}}{d{t'^{2}}}+\gamma'\frac{d{x'_l}}{d{t'}}
={\tilde{n}_e}^3 e^{-\kappa /{\tilde{n}_e}} (2+ \frac{2\kappa}
{\tilde{n}_e} + \frac{\kappa^2}{{\tilde{n}_e}^2})\ \times \\
(x'_{l+1}+x'_{l-1}-2x'_l)+ F_0^xe^{-i \omega t} \delta_{l,N}
\end{split}
\end{equation}
\begin{equation}
\begin{split}
\frac{d^{2}{y'_l}}{d{t'^{2}}}+\gamma'\frac{d{y'_l}}{d{t'}}
=-{\tilde{n}_e}^3 e^{-\kappa /{\tilde{n}_e}}(1 +
\frac{j\kappa}{\tilde{n}_e})\ \times \\
(y'_{l+1}+y'_{l-1}-2y'_l)-y'_l+ F_0^ye^{-i \omega
t}\delta_{l,\frac{N}{2}}
\end{split}
\end{equation}
\end{subequations}
with $l=1,2,...,N$ and $F_0^{x,y}$ the dimensionless strength of
the driving force. In order to excite the longitudinal vibrations
we have considered a force directed along $x$ and acting on one of
the extremity of the chain, while to excite the transversal modes
the force acts on the particle in the middle of the chain and with
$y$ component only, as was done experimentally in Ref [7]. Looking
for a particular solution of Eqs. (15a-b) of the form:
\begin{equation}
(x'_l,y'_l)=(A^x_l,A^y_l) e^{-i \omega t}
\end{equation}
we obtained the following set of inhomogeneous linear equations
for the displacements $A_l$:
\begin{subequations}
\begin{equation}
\begin{split}
{\beta}_1 A^x_{l-1}+({\omega}^2_{ac} + i \gamma {\omega}_{ac} -2
{\beta}_1)A^x_l+\\ \ + {\beta}_1 A^x_{l+1}-F_0^x \delta_{l,N}=0,
\end{split}
\end{equation}
\begin{equation}
\begin{split}
{\beta}_2 A^y_{l-1}+({\omega}^y_{opt} -1 + i \gamma {\omega}_{opt}
-2 {\beta}_2)A^y_l+\\ \ + {\beta}_2 A^y_{l+1}-F_0^y
\delta_{l,\frac{N}{2}}=0,
\end{split}
\end{equation}
\end{subequations}
where $\beta_1={\tilde{n}_e}^3 e^{-\kappa /{\tilde{n}_e}} (2+ 2
\kappa / \tilde{n}_e + \kappa^2 / {\tilde{n}_e}^2 )$ and
$\beta_2={\tilde{n}_e}^3 e^{-\kappa /{\tilde{n}_e}} (1+  \kappa /
\tilde{n}_e )$. The solution to these equations may easily be
obtained from Kramer's rule \cite{goldstein}:
\begin{equation}
A_l=\frac{D_l(\omega)}{D(\omega)}
\end{equation}
where $D(\omega)$ is the determinant of the coefficients of $A_l$
in Eqs. (17a-b) and $D_l(\omega)$ is the modification in
$D(\omega)$ resulting when the $l$th column is replaced by
$(F_0,0,0,...,0)$ for the longitudinal motion and
$(0,0,..,0,F_0,0,...,0,0)$ for the transverse motion respectively.
$A_l$ are complex quantities when $\gamma'\neq 0$ and the
formalism developed above allows to calculate amplitudes and
phases. The analytical expressions for $A_l$ are reported in
Appendix C.

In Fig. 5(a-b) we show the amplitudes of the displacements for the
longitudinal and transverse motion as a function of particle
position along the chain. The plots clearly show an exponential
decay. Regarding the displacements in the longitudinal modes,
there are edge effects, which disappear if the particle which is
excited is being in the center of the chain. In principle this
cannot be realized in the experiments, however except for the
first two particles the amplitudes of the displacements have the
same damping rate and the same phase angles both if the excited
particle is the one at the end of the chain and the one in the
middle. This is why for all the calculations we have considered a
force $F_0^x e^{-i \omega t}\delta_{l,N/2}$ instead of $F_0^x
e^{-i \omega t} \delta_{l,N}$ in Eq. (17a).

\begin{figure}
\begin{center}
\includegraphics[width=8.0cm]{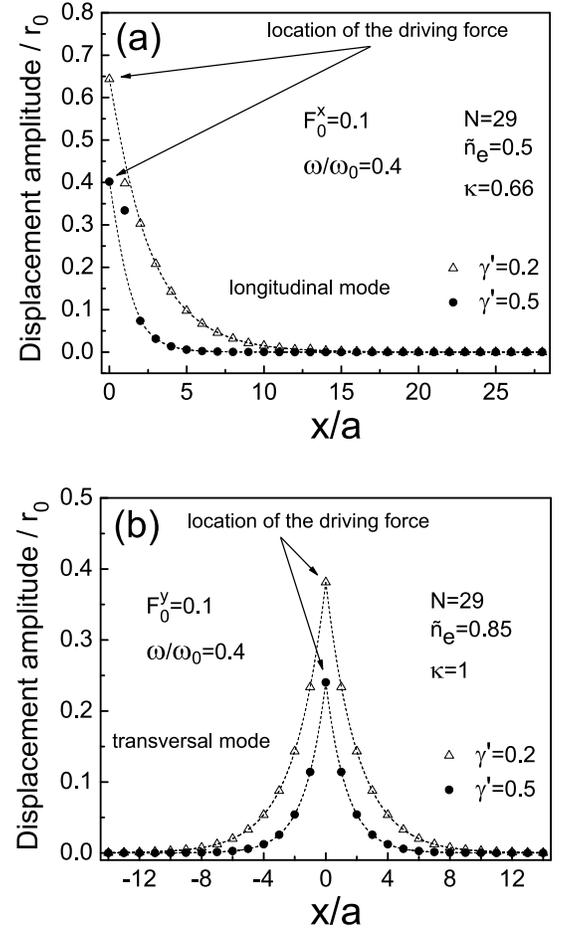}
\caption{Amplitude of the displacements of particles as a function
of the distance from the location of the driving force: (a)
displacements along the $x$-direction when the excited particle is
at the extremity of the chain, (b) displacements along $y$
direction when the excited particle is in the middle of the chain.
Exponential fits to the numerical data are shown by the dotted
curves.}
\end{center}
\end{figure}

Fitting the amplitudes to an exponential curve yields $k_i$. In
order to find the dispersion of $k_r$, we calculated the wave's
phase $\phi$ as a function of the position and fitted it to a
straight line. The definition of phase velocity, as being the
ratio between the frequency and the wave number, indeed, yields
$k_r a=\Delta \phi$. In Fig. 6 the phase angle as a function of
the distance is plotted. It is interesting to observe that $k_r$
and $k_i$ are independent of the intensity of the driving force
$F_0$, as expected in a harmonic model, and the results do not
change if instead of $F_0e^{-i \omega t}$, that is a complex
force, we consider a real force $I_0 \sin{\omega t}$, as in the
experiment; what actually plays a central role is just the driving
frequency. Optical and longitudinal waves both propagate away from
the excitation region; they are backward and forward,
respectively. It should be noticed that in a 1D chain with finite
length one should expect that only standing waves would be
allowed, the effect of gas damping is the suppression of the
reflected wave from the chain's end.

\begin{figure}
\begin{center}
\includegraphics[width=7.5cm]{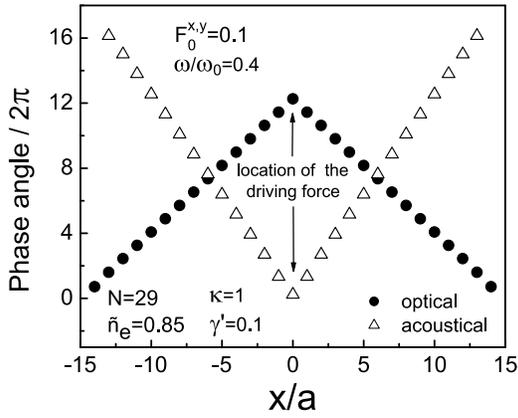}
\caption{Phase angle as a function of the distance from the
location of the driving force. The calculation are done for a
system of $N=29$ particles. Optical and longitudinal waves both
propagate away from the excitation region; they are backward and
forward, respectively.}
\end{center}
\end{figure}

It is interesting to observe that for low densities the two
calculational methods, that is the one in which the driving force
is neglected and the one in which it is explicitly taken into
account, give the same results for the dispersion curves, when
only first neighbor interactions are included. In Fig. 7 the
results of the two approaches for the optical branch are compared.
Note that outside the bond defined by the two dotted horizontal
lines the phonon mode is strongly damped.
\begin{figure}
\begin{center}
\includegraphics[width=7.7cm]{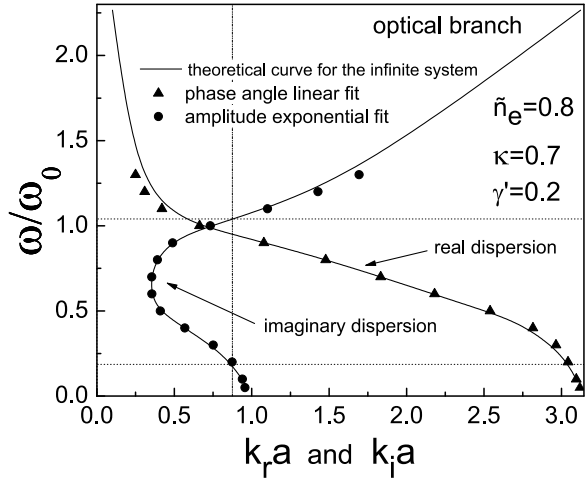}
\caption{Comparison between the standard calculation, in which the
external driving force is neglected, and the "exact" one, in which
the external driving force is explicitly considered. The region on
the left of the vertical dotted curve is the region in which the
condition $k_i a\leq \kappa/ \tilde{n}_e$ is fulfilled; the region
between the horizontal dotted curves is the region in which the
sums are convergent. Note that the two approaches give the same
results in the region of convergence.}
\end{center}
\end{figure}

Another remarkable effect, which reflects the anisotropy of the
system, is observed with increasing density: the profile of
displacements for the longitudinal mode is no longer a pure damped
exponential (see Fig. 8(a)), because reflected waves from the
chain's end start to appear, while the amplitudes for the
transversal mode are still exponentially decaying. We can infer
that the effect of damping is not simply due to friction, but also
to the external confining potential. This is confirmed by the
calculation of the amplitude profile when $\gamma'=0$, which is
reported in Fig. 8(b): even in the absence of the friction and in
the case the driving frequency is low enough, an exponential decay
of amplitudes with distance is still found.

\begin{figure}
\begin{center}
\includegraphics[width=7.5cm]{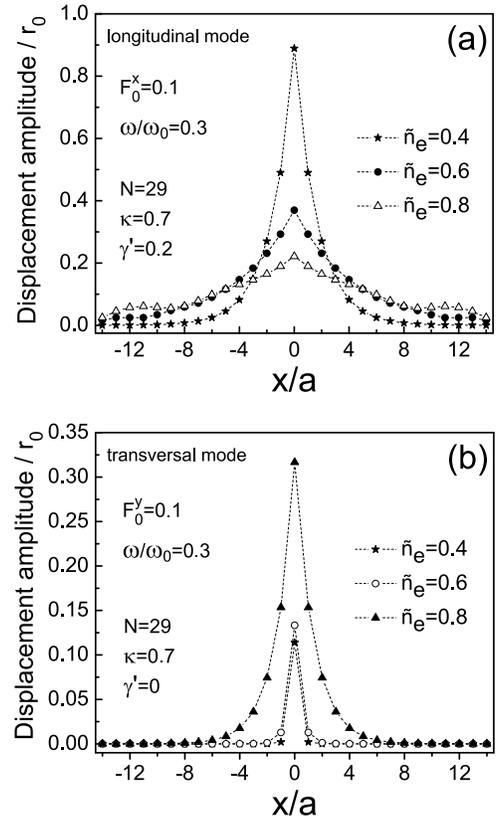}
\caption{(a) Profile of the amplitudes of displacements in the
longitudinal mode as a function of the density. For high densities
the profile is no longer a simply decaying exponential; (b)
Profile of the amplitudes of displacements in the transversal mode
as a function of the density in the case without friction. Even in
the absence of friction the profile is still a decaying
exponential.}
\end{center}
\end{figure}

When we include a perpendicular magnetic field, a coupling is
introduced between motion along $x$ and along $y$ direction. Apart
from increased mathematical complexity, the scheme developed
before is still valid. It is possible to infer the real part of
the dispersion relations from the phase angles and the imaginary
part from the amplitude of displacements. In the absence of
friction and driving force, the optical and acoustical branches
are confined in different frequency bands (see thin solid curves
in Figs. 9(a,b)), which do not cross each other and with a
prohibited gap \cite{piacente}. The optical frequencies follow the
cyclotron frequency and for very high field strength there is no
significant difference between $\omega_{opt}$ and $\omega_{c}$.
The acoustical frequencies, on the other hand, decrease with
magnetic field strength. The gap between the optical branch and
the acoustical one for large magnetic field approaches
$\omega_{c}$.

In the presence of friction and driving force, there are drastic
changes in the dispersion relations. The frequencies are no longer
confined in different bands, because the frequency of oscillation
is the one of the external force, which can be varied continously.
The results of the calculations for different intensities of the
magnetic field are shown in Figs. 9(a-b). There are no
significative differences in the behavior of the real part of the
dispersion relations with respect to the case without magnetic
field. The imaginary dispersion relations clearly show that the
waves are overdamped in the band gap, region where large values
for $k_ia$ are found. Notice that friction reduces the slope of
the acoustical branch in the small $k_ra$ region. When the curve
enters the gap region it becomes strongly damped as is clearly
seen from the inset of Figs. 9(a,b). The optical mode is for all
frequencies more strongly damped than the acoustical one.
Furthermore, the dispersion of the optical branch is strongly
modified by friction, i.e. it attains a negative dispersion for
all values of the frequency.

\begin{figure}
\begin{center}
\includegraphics[width=7.5cm]{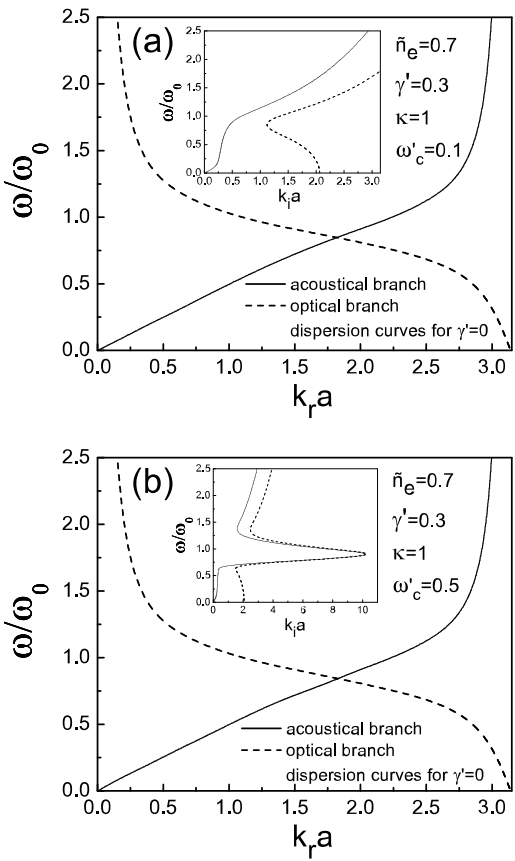}
\caption{(a) Real part of the dispersion relations for a weak
perpendicular magnetic field; (b) Real part of the dispersion
relations for a strong perpendicular magnetic field. The insets
show the imaginary part of the dispersion relations.}
\label{figurename}
\end{center}
\end{figure}

\section{Comparison with experiment}
In Figs. 10 and 11 the real and imaginary dispersion relations for
the acoustical and optical modes for the single chain
configuration are respectively presented, for different values of
the parameters.

\begin{figure}
\begin{center}
\includegraphics[width=7.8cm]{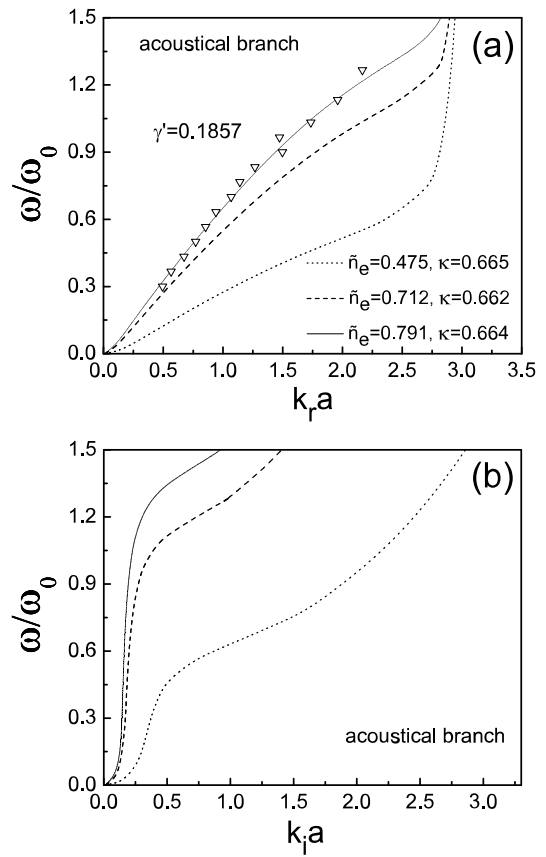}
\caption{Theoretical (curves) dispersion relations for the
acoustical mode and comparison with experimental data (symbols)
from Ref. [7] for the one chain structure: (a) real part of the
acoustical branch; (b) imaginary part of the acoustical branch.}
\end{center}
\end{figure}

The calculated dispersion relations are compared with experimental
data of Ref. \cite{goree}. The experimental data are in good
agreement with the theoretical calculations, although the system
realized in laboratory is slightly different from the one
investigated in the theory. As a matter of fact, in the experiment
the interparticle spacing was not uniform: it was 15\% smaller in
the center than in the chain's end. Due to the strong damping this
density gradient is not very important in the considered forced
oscillation. As in the case without friction, the optical mode has
negative dispersion, while the longitudinal one has positive
dispersion. The dispersion depends on density, and therefore on
interparticle spacing. For the acoustical mode the frequencies of
vibrations increase with decreasing interparticle distance, while
for optical vibration the frequencies decrease with increasing
densities. Furthermore, for low $\widetilde{n}_e$ the exponential
decay is stronger in both cases, which implies a highly damped
wave. These findings can be easily explained because for smaller
interparticle distance the interaction forces are larger, or in
other words for low densities the interaction between the
particles is rather weak and consequently the effect of a local
perturbation is less disruptive for the other particles. From Fig.
11(b) it is seen that the optical mode is mostly constrained to a
central frequency band. Comparing the optical branches in the
absence of friction (see Fig. 7 of Ref. \cite{piacente}) with the
one in the presence of gas damping, it is observed that with
damping the wave propagates beyond the frequency band allowed in
the absence of damping. For $k_r a = \pi$, $\omega_{opt}$ is
always equal to zero when $\gamma'\neq 0$ independent of the
experimental parameters. This means that in the presence of
damping, the softening of the optical mode does no longer signal a
structural phase transition from a single chain structure to a
double chain structure.

\begin{figure}
\begin{center}
\includegraphics[width=7.5cm]{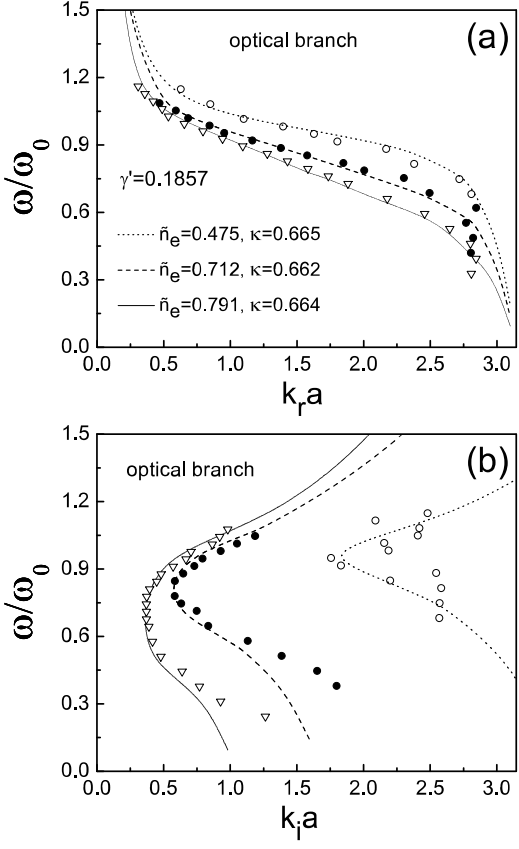}
\caption{The same as Fig. 10 but now for the optical branch.}
\end{center}
\end{figure}

It is, however, interesting to study the behavior of the
dispersion relations when the density approaches the critical
value for which the continuous structural transition from the one
chain to the two chain structure occurs. The results are shown in
Fig. 12.
\begin{figure}
\begin{center}
\includegraphics[width=7.5cm]{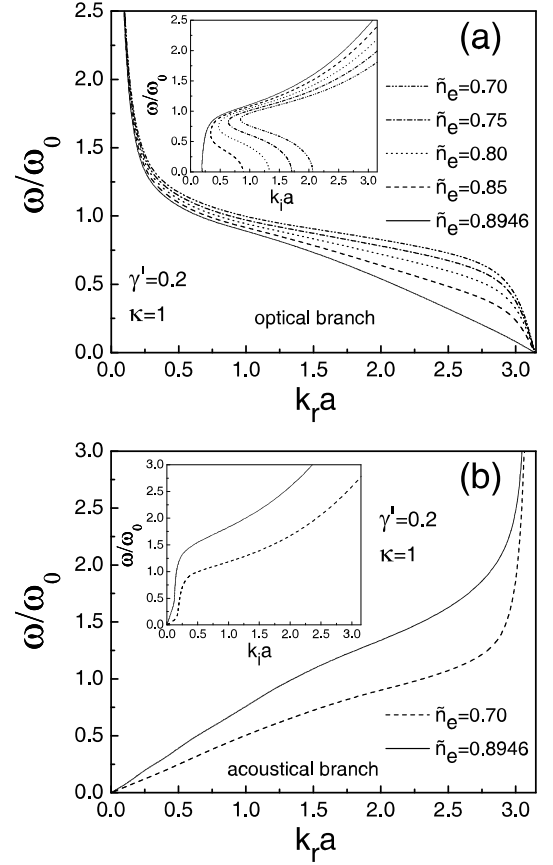}
\caption{Dispersion relations for different values of the density
approaching the critical values for which the zig-zag transition
from 1 chain structure to 2 chain structure is observed: (a)
optical branches; (b) acoustical branches. In the insets the
imaginary parts of the dispersion relations are plotted.}
\end{center}
\end{figure}
The optical branch softens when approaching the critical density,
while the acoustical branch is hardened. Notice that at the phase
transition point: i) the real part of the optical dispersion
becomes linear for $k_ra>1.5$, ii) there is a drastic change of
slope in the optical imaginary dispersion, and iii) the optical
mode becomes less damped. The real and imaginary acoustical
dispersions are less strongly influenced near the zig-zag
transition. This can be easily explained by the fact that the
zig-zag transition, which is responsible of the splitting of the
chain, acts in the $y$-direction. Therefore, signature of the
zig-zag transition are more easily detected in the optical phonon
mode.

\section{Forced oscillations in a multi-chain structure}

In Fig. 13 and Fig. 14 the real and imaginary part of the
dispersion relations for the forced oscillations of the 2 and 3
chain configurations are reported respectively. We used the
approach given in the first part of Sec. IV. Therefore, the
dispersion relations are in Fig. 13 and 14 only given in that part
of the Brillouin zone, where the sums in Eqs. (14a-b) are
convergent.

From Fig. 13(a) and Fig. 14(a) it is evident that there is a
remarkable difference in the optical branches between the single
chain and the 2 and 3 chain structures. In the first case the
optical mode has negative dispersion as stated before, while for
the 2 and 3 chain structures the optical frequencies do not
exhibit a monotonic behavior. Such feature can be attributed to
the fact that for the single chain configuration in the case of
the transversal mode the restoring force is only due to the
parabolic confining potential, while in the multi-chain
configuration the restoring force depends both on the external
confinement and on the particle repulsion.

\begin{figure}
\begin{center}
\includegraphics[width=7.5cm]{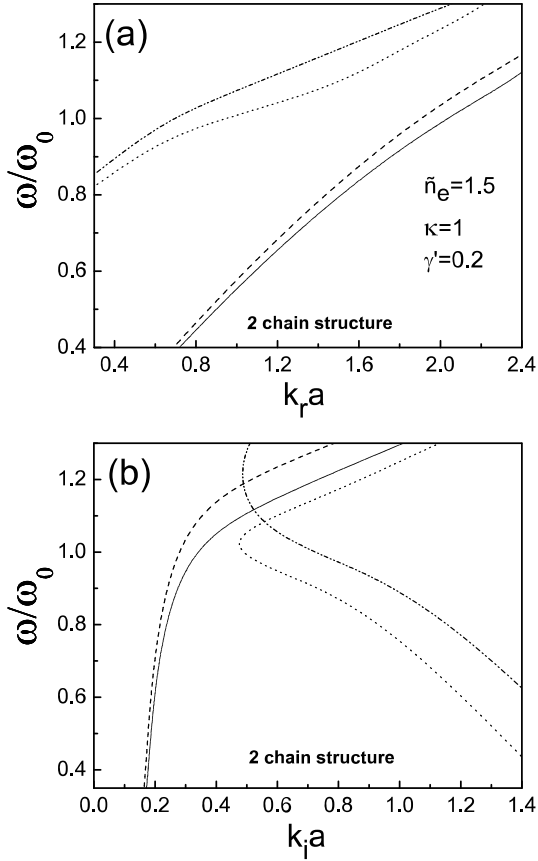}
\caption{Theoretical dispersion relations for the 2 chain
structure: (a) real part, (b) imaginary part.}
\end{center}
\end{figure}
Fig. 13(b) and Fig. 14(b) exhibit some similarities with the
single chain case: i) for the acoustical modes the damping is an
increasing function of the driving frequency, ii)the optical modes
are mostly constrained to a frequency band, and iii) the optical
modes are more strongly damped. The approach used for the
calculation of the dispersion relations for the multi-chain is the
same considered in Sec. IV for an infinite number of particles,
this is why in Figs. 13 and 14 the dispersion relations are not
presented in whole first Brillouin zone, but only in that range of
the frequency corresponding to convergent sums.
\begin{figure}
\begin{center}
\includegraphics[width=7.5cm]{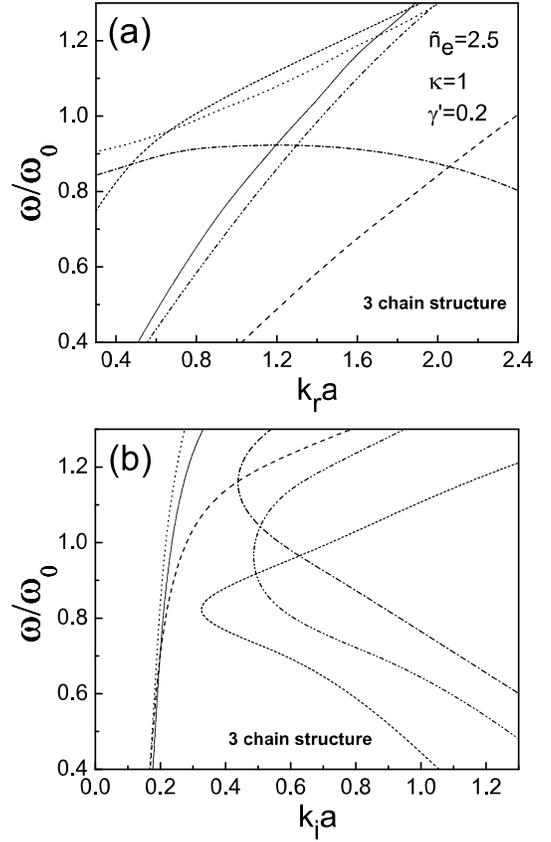}
\caption{Theoretical dispersion relations for the 3 chain
structure: (a) real part, (b) imaginary part.}
\end{center}
\end{figure}

\section{Conclusion}

The ground state and the normal modes of a Q1D multi-chain system
can be studied experimentally in a dusty plasma, where $\mu$m
sized particles are externally confined by electric fields in the
sheath above the lower electrode. The sheath conforms to the shape
of the electrode, so building up an electrode with a groove-shaped
depression in one direction allows the realization of a parabolic
confining potential and, as a consequence, the formation of a
chain-like crystal in that direction.

We investigated the structural properties and the normal modes of
such a classical Q1D system of particles interacting through a
Yukawa-type potential. The structural transitions are of first
(primarily) and second order. The normal modes of the system were
calculated first neglecting the effects of dissipation induced by
gas drag and then considering explicitly the presence of friction.
The normal modes consist of longitudinal (acoustical modes) and
transversal (optical modes). The number of acoustical branches is
equal to the number of optical branches and is equal to number of
chains in the system. In the presence of friction, the free
oscillations of the system are exponentially damped in time. The
effect of a constant magnetic field on the dispersion relations
was investigated and we found that the acoustical and optical
branches no longer cross.

Particular attention was paid to the case of forced oscillations
induced by an external driving force, as was investigated in the
experiments. We found that earlier approaches to calculate the
phonon dispersion relations are no longer valid. Our theoretical
results are compared with experimental data and a remarkably good
agreement between theory and experiment was found.

Finally, we made predictions for the single chain dispersion
relations in the presence of a perpendicular magnetic field and
for the multi-chain dispersion relations when the modes are
excited by an external driving force. We found some substantial
differences as well as some similarities in the dispersion
relations between the single and multi-chain structures.

\section*{Acknowledgment}
This work was supported in part by the European Community's Human
Potential Programme under contract HPRN-CT-2000-00157 ``Surface
Electrons'', the Flemish Science Foundation (FWO-Vl), Belgian
Science Policy, the GOA and the visitors programme of the Max
Planck Institute. We are very grateful to Prof. J. Goree for
providing us with the experimental data on the transversal mode of
the single chain structure.
\onecolumngrid
\appendix
\section{}
The equations of motion for particles in the three-chain
configuration in the presence of friction are in matrix form:
\begin{equation}
\begin{pmatrix} \ddot{x'}_n^{(1)} \\
\ddot{y'}_n^{(1)}\\
\ddot{x'}_n^{(2)}\\
\ddot{y'}_n^{(2)}\\
\ddot{x'}_n^{(3)}\\
\ddot{y'}_n^{(3)}
\end{pmatrix} =
\gamma'\begin{pmatrix} \dot{x'}_n^{(1)}\\
\dot{y'}_n^{(1)}\\
\dot{x'}_n^{(2)}\\
\dot{y'}_n^{(2)}\\
\dot{x'}_n^{(3)}\\
\dot{y'}_n^{(3)}
\end{pmatrix}+
\begin{pmatrix}  -B_1 & 0 & -B_3 & 0 & -B_5 & 0 \\
0 & -B_2 & 0 & -B_4 & 0 & -B_6 \\
-B_3 & 0 &  -B_1 & 0 & -B_3 & 0 \\
0 & -B_4 & 0 & - B_2 & 0 & -B_4 \\
-B_5 & 0 & -B_3 & 0 &  -B_1 & 0 \\
0 & -B_6 & 0 & -B_4 & 0 &  -B_2\end{pmatrix}
\begin{pmatrix} {x'_j}^{(1)} \\
{y'_j}^{(1)}\\
{x'_j}^{(2)}\\
{y'_j}^{(2)}\\
{x'_j}^{(3)}\\
{y'_j}^{(3)}\\
\end{pmatrix}
\end{equation}
\noindent where the superscript labels the row which the particle
is sitting in. The coefficients are:
\begin{eqnarray}
\nonumber B_1 &=& \frac {{\tilde{n}_e}^3}{54} \sum_{j} \frac{e^{-3
j \kappa /\tilde{n}_e}}{ j^3} \Bigg[ 2+ 6 \frac{j \kappa
}{\tilde{n}_e} + 9\frac{ j^2 {\kappa}^2}{{\tilde{n}_e}^2} \Bigg], \\
\nonumber B_2 &=& -\frac {{\tilde{n}_e}^3}{54}\sum_{j} \frac{e^{-3
j \kappa /\tilde{n}_e}}{ j^3}
\Big(1 + 3 \frac{j \kappa }{\tilde{n}_e}\Big ), \\
\nonumber B_3 &=& \frac {{\tilde{n}_e}^3}{54}\sum_{j} \frac{e^{-3
\kappa r_{12} /\tilde{n}_e}} {r_{12}^5} \Bigg [ \Big(
j+\frac{1}{2} \Big)^2 \Bigg (\frac{9 \kappa r_{12}}{\tilde{n}_e} +
\frac{9 {\kappa}^2 r_{12}^2} {{\tilde{n}_e}^2}+ 3 \Bigg )
-r_{12}^2 \Bigg(1 + \frac{3 \kappa r_{12}}{\tilde{n}_e} \Bigg)
 \Bigg ], \\
\nonumber B_4 &=& \frac {{\tilde{n}_e}^3}{54}\sum_{j} \frac{e^{-3
\kappa r_{12} /\tilde{n}_e}} {r_{12}^5} \Bigg [ {c_3}^2 \Bigg (
\frac{9 \kappa r_{12}}{\tilde{n}_e} + \frac{9 {\kappa}^2
r_{12}^2}{{\tilde{n}_e}^2}+ 3 \Bigg) - r_{12}^2 \Bigg (1 + \frac{3
\kappa r_{12}}{\tilde{n}_e} \Bigg )\Bigg ]
 , \\
\nonumber B_5 &=& \frac {{\tilde{n}_e}^3}{54}\sum_{j} \frac{e^{-3
\kappa r_{13} /\tilde{n}_e}} {r_{13}^3} \Bigg ( \frac{9 \kappa
r_{13}}{\tilde{n}_e} + \frac{9 {\kappa}^2
r_{13}^2}{{\tilde{n}_e}^2}+ 3\Bigg),\\
\nonumber B_6 &=&\frac {{\tilde{n}_e}^3}{54}\sum_{j} \frac{e^{-3
\kappa r_{13} /\tilde{n}_e}} {r_{13}^5} \Bigg [4{c_3}^2 \Bigg (
\frac{9 \kappa r_{13}}{\tilde{n}_e} + \frac{9 {\kappa}^2
r_{13}^2}{{\tilde{n}_e}^2}+ 3 \Bigg) - r_{13}^2\Bigg(1 + \frac{3
\kappa r_{13}}{\tilde{n}_e}\Bigg)\Bigg ],
\end{eqnarray}
\noindent where $r_{12}=\sqrt{(j+1/2)^2+{c_3}^2}$ and
$r_{13}=\sqrt{j^2+{4c_3}^2}$.

The equations of motion for the 2 chain structure can be obtained
by the 4 $\times$ 4 submatrixes which are included in the top left
part of the matrix in Eq. (A1): the coefficients involved in this
case are $B_1$, $B_2$, $B_3$ and $B_4$, with the substitution
${\tilde{n}_e}/3 \rightarrow \tilde{n}_e /2$.

In the presence of a constant magnetic field $\vec{B}=(0,0,B)$ the
equations of motions for the two and three chain structure are
easily obtained from the equations of motion with $B=0$, adding
the coupling terms $\dot{y'}_n^{(i)} \omega'_c$ and
$-\dot{x'}_n^{(i)} \omega'_c$ to the equations for $x$ and $y$
motion respectively, for particles sitting in the $i^{th}$ row.

\noindent Obviously, the case without gas drag is immediately
recovered setting $\gamma'=0$.

\section{}

The eigenfrequencies in the 3 chain configuration are:
\begin{eqnarray}
\nonumber \omega^{(1)}_{ac}&=&\sqrt{B_1-B_5-\gamma^2/4}-i\gamma/2,  \\
\nonumber \omega^{(1)}_{opt}&=&\sqrt{1+B_2-B_6-\gamma^2/4}-i\gamma/2,\\
\nonumber \omega^{(2)}_{ac}&=&\sqrt{B_1+B_5/2+\sqrt{B_5^2+8B_3^2}/2-\gamma^2/4}-i\gamma/2,  \\
\nonumber \omega^{(2)}_{opt}&=&\sqrt{1+B_2+B_6/2+\sqrt{B_6^2+8B_4^2}/2-\gamma^2/4}-i\gamma/2, \\
\nonumber \omega^{(3)}_{ac}&=&\sqrt{B_1+B_5/2-\sqrt{B_5^2+8B_3^2}/2-\gamma^2/4}-i\gamma/2,  \\
\nonumber
 \omega^{(3)}_{opt}&=&\sqrt{1+B_2+B_6/2-\sqrt{B_6^2+8B_4^2}/2-\gamma^2/4}-i\gamma/2,
\end{eqnarray}
\noindent where the coefficients $B_n$ are the same as in Appendix
A.
\noindent In the 2 chain configuration the eigenfrequencies
are:
\begin{eqnarray}
\nonumber \omega^{(1)}_{ac}&=&\sqrt{B_1+B_3-\gamma^2/4}-i\gamma/2,  \\
\nonumber \omega^{(1)}_{opt}&=&\sqrt{B_2+B_4-\gamma^2/4}-i\gamma/2,\\
\nonumber \omega^{(2)}_{ac}&=&\sqrt{1+B_1-B_3-\gamma^2/4}-i\gamma/2,  \\
\nonumber
\omega^{(2)}_{opt}&=&\sqrt{1+B_2-B_4-\gamma^2/4}-i\gamma/2.
\end{eqnarray}
\noindent In this case the coefficients $B_n$ are obtained from
the coefficients in Appendix A with the substitution
${\tilde{n}_e}/3 \rightarrow \tilde{n}_e /2$.

\section{}

We present the analytical expressions for the displacements
 $ A^{x,y}_l $ calculated from the Kramer's rule (Eq. (42)) in the case
of a system of $ N=29 $ particles. The central particle at which
the driving force is acting is labelled with $N=0$. For reasons of
symmetry $ A^{x,y}_{-l}=A^{x,y}_l $.
\begin{eqnarray}
\nonumber A^{x,y}_0 &=& (a^2-b^2)(-b^2-ba+a^2)(-b^2+ba+a^2)(b^4+4
b^3a-4b^2a^2-ba^3+a^4)(b^4-4 b^3 a-4 b^2 a^2+ba^3+a^4)C/D ;\\
 \nonumber
A^{x,y}_1 &=& -ab(-b^3-2b^2a+ba^2+a^3(b^3-2b^2a-ba^2+a^3)
(-7b^6+14b^4a^2-7b^2a^4+a^6)C/D;\\
 \nonumber
A^{x,y}_2 &=& ab^2(-a^6+a^5b+5a^4b^2-4a^3b^3-6a^2b^4+3ab^5+b^6)
(-a^6-a^5b+5a^4b^2+4a^3b^3-6a^2b^4-3ab^5+b^6)C/D;\\
 \nonumber
A^{x,y}_3 &=&
-ab^3(-3b^2+a^2)(a^2-b^2)(-2b^2+a^2)(b^4-4b^2a^2+a^4)C/D;\\
\nonumber A^{x,y}_4 &=& -b^4(-a^5+a^4b+4a^3b^2-3a^2b^3-3ab^4+b^5)
(a^5+a^4b-4a^3b^2-3a^2b^3+3ab^4+b^5)C/D;\\
\nonumber
 A^{x,y}_5 &=& -ab^5(-a^2-ba+b^2)(-a^2+ba+b^2)(a^4-5a^2b^2+5b^4)C/D;\\
\nonumber A^{x,y}_6 &=&
b^6(a^2-b^2)(-b^3-3b^2a+a^3)(b^3-3b^2a+a^3)C/D;\\
\nonumber A^{x,y}_7 &=& -ab^7(-2b^2+a^2)(2b^4-4b^2a^2+a^4)C/D;\\
\nonumber A^{x,y}_8 &=&
b^8(b^3-2b^2a-ba^2+a^3)(-b^3-2b^2a+ba^2+a^3)C/D;\\
\nonumber A^{x,y}_9 &=& -ab^9(-3b^2+a^2)(a^2-b^2)C/D;\\
\nonumber
 A^{x,y}_{10} &=& b^{10}(-b^2+ab+a^2)(-b^2-ab+a^2)C/D;\\
\nonumber
 A^{x,y}_{11} &=& ab^{11}(2b^2-a^2)C/D;\\
\nonumber A^{x,y}_{12}&=& b^{12}(a^2-b^2)C/D\\
\nonumber A^{x,y}_{13} &=& -ab^{13}C/D;\\
\nonumber A^{x,y}_{14} &=& -b^{14}/D,
\end{eqnarray}
\noindent where $D={a(-3 b^2+a^2)(5 b^4-5 b^2 a^2+a^4)(b^8-8 b^6
a^2+14 b^4 a^4-7 b^2 a^6+a^8)}$.

\noindent In the case of displacements along the $x$-direction:
$a= \omega'^2+i\gamma' \omega'-2b$, $b=({\tilde{n}_e}^3/2)
\exp(-\kappa/n)(2+2\kappa/{\tilde{n}_e}+\kappa^2/{\tilde{n}_e}^2)$
and $C=F_0^x$, while in the case of displacements along the
$y$-direction: $a= \omega'^2 - 1 + i\gamma' \omega'-2b$,
$b=-({\tilde{n}_e}^3/2)\exp(-\kappa/\tilde{n}_e)(1+\kappa /
\tilde{n}_e)$ and $C=F_0^y$.

\twocolumngrid

\end{document}